\documentclass[journal]{IEEEtran}

\usepackage{epsfig,amsmath,amssymb,epsf,amsthm,scalefnt,multirow,subfig}
\usepackage{xcolor}
\usepackage{float}
\usepackage{cite}
\usepackage{setspace}
\newcommand{\argmax}{\mathop{\mathrm{argmax}}}
\newcommand{\argmin}{\mathop{\mathrm{argmin}}}

\newtheorem{lemma}{Lemma}

\newtheorem{corollary}{Corollary}

  \def\cC{{\mathcal{C}}}

 \def\cN{{\mathcal{N}}}  \def\cP{{\mathcal{P}}}
   
   \def\cX{{\mathcal{X}}}

\def\argmin{\mathop{\mathrm{argmin}}}
\def\argmax{\mathop{\mathrm{argmax}}}
\def\diag{\mathop{\mathrm{diag}}}

\def\trace{\mathop{\mathrm{tr}}}

\def\b0{{\pmb{0}}} 

  \def\bg{{\mathbf{g}}} \def\bh{{\mathbf{h}}}
   
 \def\bn{{\mathbf{n}}}  
 \def\br{{\mathbf{r}}}  
  \def\bw{{\mathbf{w}}} \def\bx{{\mathbf{x}}}
\def\by{{\mathbf{y}}}   

\def\bA{{\mathbf{A}}} \def\bB{{\mathbf{B}}} \def\bC{{\mathbf{C}}} 
 \def\bF{{\mathbf{F}}}  
\def\bI{{\mathbf{I}}}  \def\bK{{\mathbf{K}}} 
   \def\bP{{\mathbf{P}}}
 \def\bR{{\mathbf{R}}}  
\def\bU{{\mathbf{U}}} \def\bV{{\mathbf{V}}}  \def\bX{{\mathbf{X}}}
\def\bY{{\mathbf{Y}}}

\begin{document}

% paper title
\title{Downlink Training Techniques for FDD Massive MIMO Systems: Open-Loop and Closed-Loop Training with Memory}

 %author names and IEEE memberships
\author{Junil Choi$^*$, David J. Love, and Patrick Bidigare\\
\thanks{Copyright (c) 2013 IEEE. Personal use of this material is permitted. However, permission to use this material for any other purposes must be obtained from the IEEE by sending a request to pubs-permissions@ieee.org.}
\thanks{Junil Choi and David Love are with the School of Electrical and Computer Engineering, Purdue University, West Lafayette, IN (e-mail: choi215@purdue.edu, djlove@ecn.purdue.edu).}
\thanks{Patrick Bidigare is with Raytheon BBN Technologies, Arlington, VA (e-mail: bidigare@ieee.org).}
\thanks{This paper was presented in part at the IEEE Conference on Information Sciences and Systems, Johns Hopkins University, 2013 \cite{closedloop_training}.}}
% The paper headers
%\markboth{Draft}%
%{Paper Outline}

% make the title area
\maketitle

%\squeezeup\squeezeup\squeezeup\squeezeup
\begin{abstract}
The concept of deploying a large number of antennas at the base station, often called massive multiple-input multiple-output (MIMO), has drawn considerable interest because of its potential ability to revolutionize current wireless communication systems.  Most literature on massive MIMO systems assumes time division duplexing (TDD), although frequency division duplexing (FDD) dominates current cellular systems.  Due to the large number of transmit antennas at the base station, currently standardized approaches would require a large percentage of the precious downlink and uplink resources in FDD massive MIMO be used for training signal transmissions and channel state information (CSI) feedback.  To reduce the overhead of the downlink training phase, we propose practical open-loop and closed-loop training frameworks in this paper.  We assume the base station and the user share a common set of training signals in advance.  In open-loop training, the base station transmits training signals in a round-robin manner, and the user successively estimates the current channel using long-term channel statistics such as temporal and spatial correlations and previous channel estimates.  In closed-loop training, the user feeds back the best training signal to be sent in the future based on channel prediction and the previously received training signals.  With a small amount of feedback from the user to the base station, closed-loop training offers better performance in the data communication phase, especially when the signal-to-noise ratio is low, the number of transmit antennas is large, or prior channel estimates are not accurate at the beginning of the communication setup, all of which would be mostly beneficial for massive MIMO systems.
\end{abstract}

\section{Introduction}
%Multiple-input multiple-output (MIMO) systems have been thoroughly studied over the past two decades, and the advantages of MIMO systems are now well known.  In point-to-point communication scenarios, MIMO systems can increase the received signal-to-noise-ratio (SNR) by using transmit beamforming \cite{bf1,bf2,bf3} or space-time block codes \cite{stbc1,stbc2,stbc3} and multiplexing gain by sending multiple data streams using the same time and frequency resource block \cite{Tse}.  Moreover, by supporting multiple users with the same resources in time and frequency, multi-user (MU) MIMO systems can theoretically boost the system throughput significantly \cite{mu_mimo2}.

Recently, massive multiple-input multiple-output (MIMO) systems that employ a very large number of antennas, e.g., tens or hundreds of antennas, at the base station have drawn considerable interest from both academia and industry.  Theoretically, massive MIMO systems can almost perfectly alleviate the inter-user interference that occurs in downlink and uplink multiuser MIMO (MU-MIMO) systems with a simple linear precoder and receive combiner \cite{massive_mimo1}.  Moreover, the transmission power of each antenna can be made arbitrarily small as the number of antennas increases \cite{massive_mimo2}, which makes it possible to implement massive MIMO systems with cheap, linear power amplifiers.  Massive MIMO can also increase network throughput by supporting a large number of users simultaneously \cite{massive_mimo4}.  We refer to \cite{massive_mimo3} and references therein for more about the concept of massive MIMO systems.

Note that the optimal benefits of MIMO and massive MIMO system can be achieved only when the base station and the user (or multiple users in MU-MIMO systems) both know the channel state information (CSI) between the two perfectly.  However, it is impossible for the base station and the user to know the CSI perfectly in practice.  Instead, the user acquires the CSI through a training phase in which the base station transmits training signals that are known at the user a priori.  To provide transmit-side CSI, the base station can learn the CSI from limited feedback in frequency division duplexing (FDD) \cite{jsac_limited_feedback} or leverage channel reciprocity in time division duplexing (TDD) \cite{massive_mimo1}.

Most of the massive MIMO research assumes TDD systems that rely on channel reciprocity for the base station to acquire CSI.  Ideally, pilot contamination, which is caused by using non-orthogonal uplink pilot signals in neighbouring cells, is the only factor that limits TDD massive MIMO system performance \cite{massive_mimo1,massive_mimo5}.  Some works that mitigate pilot contamination have been proposed recently \cite{Yin:2013,Muller:2013}.  However, in practice, there are other system imperfections that limit the performance of TDD massive MIMO systems.  Because of calibration error in the downlink/uplink RF chains, the downlink channel estimated by the uplink channel using channel reciprocity may not be accurate \cite{tdd_cali}.  Hardware impairments also can limit the performance \cite{Pitarokoilis:2013,Emil:2013}.  Moreover, the user is not able to learn the instantaneous downlink channel (because there is no downlink training for CSI estimation in TDD massive MIMO) \cite{massive_mimo5}, which might cause a significant error in data decoding at the user.  In addition, FDD dominates current wireless cellular systems.  Thus, we focus on a downlink training framework for FDD massive MIMO systems in this paper.

Many papers have been dedicated to deriving the optimal training signals for \textit{open-loop/single-shot} training frameworks and verifying their channel estimation performance in FDD MIMO systems \cite{training1,training2,training3,training4,training5,song}.  Open-loop training means that there is no feedback information about the preferable training signal, and single-shot training refers to the case when the user estimates the channel only based on the current received training signal and discards the past received training signals.  In open-loop/single-shot training, it was shown in \cite{training1} that training signals should be orthogonal to each other, and the optimal training length in time should be the same as the number of transmit antennas in uncorrelated Rayleigh fading channels.  When channels are spatially correlated and the base station knows the correlation statistic perfectly, \cite{training2} and \cite{training5} showed that the optimal training dimension can be reduced when the number of statistically dominant subspaces is smaller than the number of transmit antennas.  In temporally correlated channels, a Kalman filter or particle filter can be used at the user to track the channel variation between the training signal intervals \cite{ali,particle1}.

The amount of temporal overhead for downlink training has been assumed negligible in past MIMO scenarios because past systems used small numbers of transmit antennas.  However, in FDD massive MIMO systems, the overhead of the training duration could overwhelm the precious downlink resources due to the large number of transmit antennas.  Therefore, we propose practical open-loop and closed-loop training approaches with successive channel estimation for FDD massive MIMO in order to reduce the overhead of the downlink training phase.

We consider practical MIMO channels that are correlated in time \textit{and} space.  Moreover, we assume that the long-term channel statistics are known only to the user.  This assumption is different from \cite{training2,training5,song} that assume perfect knowledge of the spatial correlation at the base station.  Having spatial correlation knowledge at the base station may not be practical for FDD massive MIMO systems because the user would have to explicitly feed back the knowledge of the spatial correlation matrix to the base station.  Since the number of entries of the spatial correlation matrix grows quadratically with the number of transmit antennas, feedback of the spatial correlation matrix might not be acceptable in massive MIMO systems.\footnote{When statistical reciprocity is available, it is also possible to estimate the downlink spatial correlation matrix by the uplink correlation matrix to sidestep this problem \cite{up_down_convert}.}  The spatial and temporal correlation vary in time in practice, even though they are long-term channel statistics, which makes it even harder for the base station to acquire such statistics.  Thus, we assume \textit{the base station does not have any knowledge of those statistics} throughout the paper.

The contributions of this paper are summarized as follows.

\noindent $\bullet$ We first explain the limitations of single-shot training, which only relies on the most recently received training signal to estimate the channel.  The analysis shows that the average received signal-to-noise ratio (SNR) quickly saturates to a certain level as the number of transmit antennas gets large with a fixed training length (that is less than the number of transmit antennas), and this SNR ceiling could preclude its use in massive MIMO systems.

\noindent $\bullet$ We propose open-loop and closed-loop training frameworks with memory to effectively alleviate the SNR ceiling effect.  Although the ceiling effect cannot be perfectly eliminated with a fixed training length, the proposed training frameworks can significantly increase the ceiling level.  We assume the base station and the user share a common set of training signals where each training signal has a much lower rank than the number of transmit antennas.  In open-loop training, the base station transmits training signals in a round-robin manner, and the user predicts/estimates the channel based on previous channel estimates.  Thus, the proposed framework can be considered \textit{open-loop training with memory}.  With this approach, we can reduce the number of training channels needed to acquire a \textit{good} channel estimate to a reasonable range even with a large number of transmit antennas.

In \textit{closed-loop training with memory}, which had initial results presented in \cite{closedloop_training} and was studied for the stationary channel in \cite{duly_cl_training}, the user selects the best training signal based on the prior knowledge of the channel and previously received training signals.  The user feeds back the index of the selected training signal and the base station relies on the fed back information for the next training phase.  This framework is considerably different from current wireless systems that use pre-determined training signals in time and frequency \cite{lte,Yin:2013,Dai:2013}.  By allowing a small amount of feedback, we can further improve channel estimation performance with less training overhead.  We develop two objective functions to select the training signal at the user: 1) minimizing mean squared error (MSE) and 2) maximizing the average received SNR for the data communication phase.  Numerical studies show that the second approach can improve the received SNR when the number of transmit antennas is moderately large.  We also develop an effective way of designing the set of training signals used for the proposed training frameworks.

\noindent $\bullet$ We identify preferable channel conditions and system parameters for closed-loop training with memory.  The performance gain of closed-loop training becomes larger when 1) the SNR is low, 2) the number of transmit antennas is large relative to the length of the training phase, or 3) the prior channel estimate is not accurate at the beginning of the communication setup, all of which could be commonly true for massive MIMO systems. Simulation results confirm these analyses.

The remainder of the paper is organized as follows.  We present the system model we consider in Section \ref{sys_model}.  In Section \ref{sat_effect}, we first explain the concept of the conventional single-shot training and derive the structure of the optimal training signal for single-shot training assuming full feedback from the user to the base station.  Then, we show the limitation of single-shot training for massive MIMO systems.  We propose open-loop and closed-loop training frameworks with memory for massive MIMO systems in Section \ref{cl_training_framework}.  Simulation results that verify the effectiveness of the proposed schemes are presented in Section \ref{sim}, and conclusions follow in Section \ref{conclusion}.

\textbf{\textit{Notations}}: Upper and lower boldface symbols are used to denote matrices and column vectors, respectively. $\bX^H$, $\bX^T$, $\bX^{-1}$, $\bX^{\frac{1}{2}}$, and $\mathrm{tr}(\bX)$ are used as the Hermitian transpose, transpose, inverse, square-root, and the trace of $\bX$, respectively.  $\bI_{k}$ is the $k \times k$ identity matrix, and $\bX_{[k:m]}$ represents the sub-matrix of $\bX$ formed by the $k$-th column to the $m$-th column, inclusively.  $\|\bX\|$ and $\|\bX\|_F$ are used as the two-norm and the Frobenius norm of a matrix $\bX$, respectively.  We let $\lambda\left(\bX\right)$ denote the vector with the eigenvalues of the matrix $\bX$ in decreasing order as its elements.  $\mathrm{mod}(a,b)$ is the remainder of $a$ when divided by $b$.  $\cC\cN(\bar{\bx},\bR)$ is used to denote the complex Gaussian random vector distribution with mean $\bar{\bx}$ and covariance $\bR$.  $\mathbb{C}^{m}$ and $\mathbb{C}^{m\times n}$ represent a set of all $m \times 1$ complex vectors and a set of all $m\times n$ complex matrices, respectively.  The expectation operation is denoted by $E[\cdot]$.

\section{System Model}\label{sys_model}
We consider an $N_t$ transmit antennas and single receive antenna multiple-input single-output (MISO) system transmitting over a block-fading channel.  Although we only consider MISO channels for simplicity, our framework can be easily extended to general MIMO channels with the vectorization approach in \cite{training2,training5}.  We assume the block-fading channel has a coherence time of $L$, which means that the channel is static for $L$ channel uses in each block and changes from block-to-block.  The input-output relation for the $\ell$-th channel use in the $i$-th fading block is given by
\begin{equation}\label{scalar_inputoutput}
y_{i}[\ell] = \bh_{i}^H\bx_{i}[\ell] + n_{i}[\ell],
\end{equation}
where $y_{i}[\ell]$ is the received signal, $\bh_{i}\in\mathbb{C}^{N_t}$ is the channel vector, $\bx_i[\ell]\in\mathbb{C}^{N_t}$ is the transmitted signal with $E[\|\bx_i[\ell]\|^2] = \rho$, and $n_i[\ell]\sim\cC\cN(0,1)$ is normalized additive white Gaussian noise at the user.

Each channel block consists of a \textit{training} phase and a \textit{data communication} phase.  We assume that the first $T< L$ channel uses and the remaining $L-T$ channel uses are dedicated for training and data communication, respectively.  We further assume that $T<N_t$ because we consider massive MIMO.  For the $i$-th fading block, the received training signals $y_{i}[\ell]$ for $\ell=0,\ldots,T-1$ can be collected into a vector form as $\by_{i,train} = \left[y_i[0]~\cdots~y_i[T-1]\right]^T$.  Then, the input-output relation in \eqref{scalar_inputoutput} can be rewritten as
\begin{equation}\label{eq_train}
\by_{i,train} = \bX_i^H \bh_i+\bn_{i,train},
\end{equation}
where $\bX_i=\left[\bx_i[0]~\cdots~\bx_i[T-1]\right]$ is the transmitted signals collected into an $N_t \times T$ matrix and $\bn_{i,train} = \left[n_i[0]~\cdots~n_i[T-1]\right]^T$.

Note that \textit{unitary training} with equal power allocation per pilot symbol which restricts $\bX_i$ such that
\begin{equation}\label{unitary_X}
  \bX_i \in \cX = \left\{\bF:\bF\in \mathbb{C}^{N_t \times T},~\bF^H\bF = \rho \bI_T\right\},
\end{equation}
is optimal in i.i.d. Rayleigh fading channels.  We also rely on unitary training throughout this paper because we assume that the base station does not have any prior knowledge of the channel statistics to adapt training signals.\footnote{If the base station knows the channel statistics, then non-unitary training with power allocation can give a better performance than unitary training in spatially correlated channels.}

During the $L-T$ data communication channel uses, we assume that the base station employs beamforming, and the transmitted signal is written as
\begin{equation}\notag
\bx_i[\ell] = \bw_i s_i[\ell],
\end{equation}
where $\bw_i$ is a beamforming vector with $\|\bw_i\|=1$ and $s_i[\ell]$ is a data symbol with  $E[\left|s_i[\ell]\right|^2]=\rho.$  With this setup, the normalized average received SNR of the $i$-th fading block at the user is
\begin{equation}\label{snr_average}
  \Gamma_i = \frac{1}{\rho}E\left[|\bh_i^H \bx_i[\ell]|^2\right]= E\left[|\bh_i^H \bw_i|^2\right].
\end{equation}

The optimal training signal $\bX_i$ is highly dependent on the channel statistics and the desired performance metric.  Aside from the few works, including \cite{training2,training5} that assume spatially correlated channels and \cite{particle1} that assumes temporally correlated channels, most research on training considers uncorrelated channels both in time and space.  In this paper, we consider a general and practical channel model, i.e., spatially \textit{and} temporally correlated channels.  We assume $\bh_i$ follows a Gauss-Markov distribution according to
\begin{align}\label{h_model}
\nonumber    \bh_0 &= \bR^{\frac{1}{2}} \bg_0,\\
    \bh_i &= \eta \bh_{i-1}+\sqrt{1-\eta^2}\bR^{\frac{1}{2}} \bg_i,\quad i\geq 1,
\end{align}
where $\bR=E\left[\bh_i\bh_i^H\right]$ is a spatial correlation matrix,\footnote{$\bR$ is closely related to the antenna spacing at the base station and the user location.  We assume that $\bR$ is fixed in time because the user location does not change much with moderate user velocities, e.g., 3-10km/h.} $\bg_i$ is an innovation process with independent and identically distributed (i.i.d.) entries distributed according to $\cC\cN(\mathbf{0},\bI_{N_t})$ for all $i$, and $0\leq \eta \leq 1$ is a temporal correlation coefficient.  We assume $\bh_0$ is independent of $\bg_i$ for all $i\geq 1$. Because the spatial correlation matrix $\bR$ is a Hermitian positive definite matrix, it can be decomposed as
$\bR=\bU\mathbf{\Lambda}\bU^H$ where $\bU$ and $\mathbf{\Lambda}=\diag\left(\left[\lambda_1,\lambda_2,\cdots,\lambda_{N_t}\right]\right)$ are the eigenvector and eigenvalue matrices of $\bR$, respectively.  We assume the $\lambda_k$'s are in decreasing order as $\lambda_1\geq \cdots \geq \lambda_{N_t}$ and $\trace\left(\bR\right)=\sum\limits_{t=1}^{N_t}\lambda_t = N_t$.  As mentioned in the introduction, we assume \textit{the base station does not have any knowledge of the channel statistics such as $\bR$ and $\eta$} throughout the paper.

\section{Single-Shot Training and the Ceiling Effect}\label{sat_effect}
In most prior work on training, the user discards the previously received training signals $\left\{\by_{k,train}\right\}_{k=0}^{i-1}$ and estimates $\bh_i$ based only on the current received training signal $\by_{i,train}$.  We first explain the conventional single-shot training framework and derive the structure of the optimal training signal $\bX_{i,\mathrm{opt}}$ for single-shot training at the $i$-th fading block assuming there is an \textit{unlimited feedback channel} for $\bX_{i,\mathrm{opt}}$ from the user to the base station ($\bX_{i,\mathrm{opt}}$ is available only at the user because the base station does not know the channel statistics).  Based on an upper bound on training performance for single-shot training using the optimal training signal, we show that deploying a large number of transmit antennas does not increase performance (i.e., the normalized average received SNR $\Gamma_i$ in \eqref{snr_average}) with $N_t$ for most practical channel conditions.

We drop the fading block index $i$ from the notation throughout this section because of the lack of dependence on the specific block during channel estimation.

\subsection{Structure of the optimal training signal of single-shot training}
We focus on minimum mean squared error (MMSE) channel estimation at the user.  Assuming $\bh$ is complex Gaussian with mean $\mathbf{0}$ and covariance $\bR$, we can derive the MMSE estimate of the channel $\bh$ given the observation $\by_{train}$ in \eqref{eq_train} as \cite{Kay}
\begin{alignat}{2}\notag
\widehat{\bh} &= E[\bh\mid\by_{train}] \\\notag&= \bR\bX\left(\bI_T+\bX^H\bR\bX\right)^{-1}\by_{train}.
\end{alignat}
This estimate $\widehat{\bh}$ is complex Gaussian with mean $\mathbf{0}$ and covariance
\begin{equation}\notag
\bR_{\widehat{\bh}} = \bR\bX\left(\bI_T +\bX^H\bR\bX\right)^{-1}\bX^H\bR.
\end{equation}
The MSE of channel estimation is given by
\begin{align}
\nonumber \mathrm{MSE}\left(\bX\right) & = \frac{1}{N_t}E\left[\|\bh - \widehat{\bh}\|^2\right] \\
 & = \frac{1}{N_t}\mathrm{tr}\left(\bR-\bR\bX \left(\bI_T + \bX^H\bR \bX \right)^{-1} \bX^H\bR\right),\label{mse_orig}
\end{align}
and MMSE estimation minimizes the MSE between $\bh$ and $\widehat{\bh}$ for a given $\bX$.

As mentioned in \eqref{unitary_X}, we assume $\bX\in\cX$.  If the base station relies on a pre-defined $\bX$ for training, we call it \textit{open-loop/single-shot training}.  If there is a feedback channel from the user to the base station to inform the best training signal for single-shot training, we call this scheme \textit{closed-loop/single-shot training}.  Then, similar to the derivation in \cite{training2,training5}, the following lemma shows the optimal structure of $\bX_{\mathrm{ss,opt}}$ that minimizes the $\mathrm{MSE}\left(\bX\right)$ in \eqref{mse_orig} for closed-loop/single-shot training with unlimited feedback.

\begin{lemma}\label{opt_X}
The optimal $N_t \times T$ ($N_t \geq T$) training signal for closed-loop/single-shot training with full feedback for $\bX_{\mathrm{ss,opt}}$ that minimizes the $\mathrm{MSE}\left(\bX\right)$ is given as
\begin{align}
\nonumber  \bX_{\mathrm{ss,opt}} &= \argmin_{\bX\in \cX}\mathrm{MSE}\left(\bX\right)\\
  & = \sqrt{\rho} \bU_{[1:T]}\label{Xopt_orig}
\end{align}
where $\bR=\bU\mathbf{\Lambda}\bU^H$.
\end{lemma}
\begin{IEEEproof}
See Appendix A.
\end{IEEEproof}

Lemma \ref{opt_X} implies we should transmit the training signal along the first $T$ dominant eigen-directions of $\bR$ to minimize the MSE.  With $\bX_{\mathrm{ss,opt}} = \sqrt{\rho} \bU_{[1:T]}$, we can derive the MSE as
\begin{align}
\nonumber  &\mathrm{MSE}\left(\bX_{\mathrm{ss,opt}}\right)\\
\nonumber &= 1 - \frac{1}{N_t}\trace\left(\bR\bX_{\mathrm{ss,opt}}\left(\bI_T +\bX_{\mathrm{ss,opt}}^H\bR\bX_{\mathrm{ss,opt}}\right)^{-1}\bX_{\mathrm{ss,opt}}^H\bR\right)\\
\nonumber &= 1 - \frac{1}{N_t}\trace\left(\left(\bI_T +\bX_{\mathrm{ss,opt}}^H\bR\bX_{\mathrm{ss,opt}}\right)^{-1}\bX_{\mathrm{ss,opt}}^H\bR^2\bX_{\mathrm{ss,opt}}\right)\\
&\stackrel{(a)}{=} 1 - \frac{1}{N_t}\sum_{t=1}^{T}\frac{\rho \lambda_t^2}{\rho\lambda_t +1}\label{tr_calc}
\end{align}
where $(a)$ follows from $\bR=\bU\mathbf{\Lambda}\bU^H$.  From \eqref{tr_calc}, we state following lemma\footnote{The result similar to Lemma 2 was already proven in Theorem 2 of \cite{training5}; however, we believe the proof in this paper is of value due to its simplicity.} and corollaries, which are intuitive.

\begin{lemma}\label{sp_help}
Let $\bR_H$ and $\bR_L$ denote two $N_t \times N_t$ spatial correlation matrices.  We assume $\lambda\left(\bR_H\right)$ majorizes $\lambda\left(\bR_L\right)$, i.e., $\lambda\left(\bR_H\right)\succ \lambda\left(\bR_L\right)$ which corresponds to the case when $\bR_H$ is more spatially correlated than $\bR_L$ \cite{training5}.  We let $\bX_{H}$ and $\bX_{L}$ denote the optimal $N_t \times T$ orthogonal single-shot training signals for channels correlated with $\bR_H$ and $\bR_L$, respectively.  Then, we have
\begin{equation*}
  \mathrm{MSE}\left(\bX_H\right) \leq \mathrm{MSE}\left(\bX_L\right).
\end{equation*}
\end{lemma}
\begin{IEEEproof}
See Appendix B.
\end{IEEEproof}
\begin{corollary}\label{coro1}
If $\rho$ and $\left\{\lambda_t\right\}_{t=1}^{T}$ are fixed and $T_1 > T_2\geq 1$, then
\begin{equation*}
  \mathrm{MSE}\left(\bX_{\mathrm{ss,opt}}(T_1)\right)<\mathrm{MSE}\left(\bX_{\mathrm{ss,opt}}(T_2)\right).
\end{equation*}
\end{corollary}
\begin{corollary}\label{coro2}
If $T$ and $\left\{\lambda_t\right\}_{t=1}^{T}$ are fixed and $\rho_1 > \rho_2 > 0$, then
\begin{equation*}
  \mathrm{MSE}\left(\bX_{\mathrm{ss,opt}}(\rho_1)\right)<\mathrm{MSE}\left(\bX_{\mathrm{ss,opt}}(\rho_2)\right).
\end{equation*}
\end{corollary}
Lemma \ref{sp_help}, Corollary \ref{coro1} and \ref{coro2} show that the channel can be estimated with lower MSE when channels are highly correlated, using more time on training, or training with higher transmit power.  Although the above statements are for the case of $\bX_{\mathrm{ss,opt}} = \sqrt{\rho} \bU_{[1:T]}$, numerical results in Section \ref{sim} show that these statements also hold for a general training signal $\bX$.

\subsection{Ceiling effect of single-shot training}\label{ceiling_effet}
We assume that the user can feed back not only $\bX_{\mathrm{ss,opt}}$ but also the estimated channel $\widehat{\bh}$  perfectly to the base station to focus only on the effect of training.  We refer to \cite{tcom_ntcq,tec} and references therein that discuss the downlink CSI quantization problem in FDD massive MIMO systems.  The base station can then set the beamforming vector to $\bw = \frac{\widehat{\bh}}{\|\widehat{\bh}\|}.$
%\begin{equation*}
%\bw = \frac{\widehat{\bh}}{\|\widehat{\bh}\|}.
%\end{equation*}
Based on $\bX_{\mathrm{ss,opt}}$ and $\bw$, we derive an upper bound of the normalized average received SNR using single-shot training in the following lemma.
\begin{lemma}\label{snr_approx}
With the training signal $\bX_{\mathrm{ss,opt}} = \sqrt{\rho} \bU_{[1:T]}$ and the beamforming vector $\bw = \frac{\widehat{\bh}}{\|\widehat{\bh}\|}$, the normalized average received SNR of single-shot training, $\Gamma_{\mathrm{ss,opt}}$, can be upper bounded as
\begin{equation}
\Gamma_{\mathrm{ss,opt}} =E\left[\left|\bh^H \frac{\widehat{\bh}}{\|\widehat{\bh}\|}\right|^2\right] \leq \sum_{t=1}^{T}\frac{\rho \lambda_t^2}{\rho\lambda_t +1} + \lambda_1 \label{gamma_upper1}
\end{equation}
where $1\leq T \leq N_t$ and $\lambda_t$ is the $t$-th dominant eigenvalue of $\bR$.
\end{lemma}
\begin{IEEEproof}
See Appendix C.
\end{IEEEproof}
The upper bound in \eqref{gamma_upper1} becomes trivial when the rank of $\bR$ is 1, i.e., $\trace\left(\bR\right)=\lambda_1 = N_t$.  However, \eqref{gamma_upper1} is a non-trivial upper bound in general.

Lemma \ref{snr_approx} shows that $\Gamma_{\mathrm{ss,opt}}$ is not a linearly increasing function of $N_t$ although the impact of $N_t$ is implicitly reflected in $\lambda_t$.  With the extreme case of i.i.d. Rayleigh fading channels where $\lambda_t=1$ for all $t$, $\Gamma_{\mathrm{ss,opt}}$ is fixed to a constant with a given $T$ and $\rho$ even when $N_t\rightarrow \infty$.  Unless the dominant eigen-directions contain most of the gain of the wireless channel, which is rarely the case even in highly correlated channels in practice, the gain of having a large number of antennas will saturate eventually.

Now we verify Lemma \ref{snr_approx} with Rayleigh fading channels which are spatial correlated with the exponential model of $\bR$ that is given as\footnote{We adopt the exponential model of $\bR$ for simulation purposes.  Other structures of $\bR$ such as a Kronecker model can be adopted as well.}
\begin{align}\label{exp_model}
\bR=
\begin{bmatrix}
  1 & a & \cdots & a^{N_t-1}\\
  a & 1 &  &  \\
  \vdots &  & \ddots &  \\
  a^{N_t-1} &  &  & 1\end{bmatrix}
\end{align}
where $0<a<1$ is a real number.  The amount of spatial correlation is controlled by $a$, i.e., a larger (smaller) value of $a$ corresponds to highly (loosely) correlated channels in space.  When $a=0$, we have i.i.d. Rayleigh fading channels.

Before showing the numerical results, we state the following corollary which use the upper bound of the maximum eigenvalue of $\bR$ of the exponential model \cite{bound_old}
\begin{equation*}
  \lambda_1 \leq \frac{1+a}{1-a}.
\end{equation*}
\begin{corollary}\label{coro3}
With the exponential model of $\bR$ in \eqref{exp_model}, $\Gamma_{\mathrm{ss,opt}}$ can be further upper bounded as
\begin{equation}\label{coro3_eq}
\Gamma_{\mathrm{ss,opt}}\leq \sum_{t=1}^{T}\frac{\rho \lambda_t^2}{\rho\lambda_t +1} + \lambda_1 < (T+1)\lambda_1\leq (T+1)\frac{1+a}{1-a}.
\end{equation}
\end{corollary}
Corollary \ref{coro3} states that the maximum $\Gamma_{\mathrm{ss,opt}}$ is a function of $T$ and $a$, not $N_t$.

\begin{figure}[t]
\centering
\includegraphics[width=0.9\columnwidth]{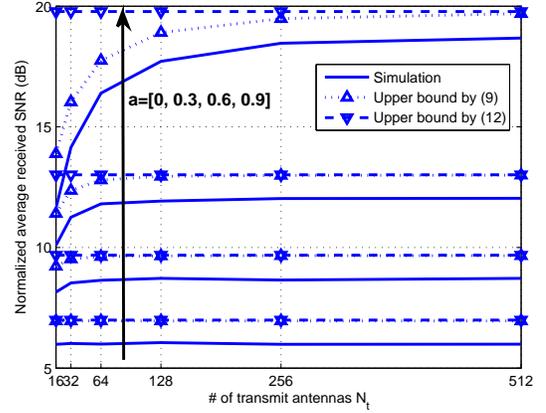}\\
\caption{Plots of $\Gamma_{\mathrm{ss,opt}}$ (in dB scale) with simulation results and the upper bounds in \eqref{gamma_upper1} and \eqref{coro3_eq} with $\rho=20$dB and $T=4$.  The ordered $a$ values by the arrow correspond with the curves moving from bottom to top.}\label{ga_opt_fig}
\end{figure}

In Fig. \ref{ga_opt_fig}, we plot $\Gamma_{\mathrm{ss,opt}}$ (in dB scale) based on simulation and the upper bounds in \eqref{gamma_upper1} and \eqref{coro3_eq}.  From the figure, we see that $\Gamma_{\mathrm{ss,opt}}$ saturates even with the optimal $\bX_{\mathrm{ss,opt}}$ and very highly correlated case of $a=0.9$.  Note that the maximum possible value of normalized average received SNR is the same as the number of transmit antennas $N_t$. We can increase $\Gamma_{\mathrm{ss,opt}}$ by using a large number of channel uses $T$ for training, but this will decrease the number of channel uses $T-L$ for the actual data communication.

The ceiling effect can be effectively reduced by exploiting the temporal channel correlation.  Although temporal correlation is present essentially for all wireless communication systems, this correlation is not widely exploited in most MIMO channel estimation and training works.  Training for massive MIMO systems should leverage temporal correlation of the channel to maximize the benefit of having a large number of antennas.

\section{Proposed Training Frameworks}\label{cl_training_framework}
In this section, we first explain \textit{open-loop training with memory} that does not require any feedback from the user to the base station.  We then propose the framework of \textit{closed-loop training with memory}.  We derive a performance upper bound of closed-loop training with memory assuming the perfect feedback of the training signal from the user to the base station.  We also present an effective way of designing the set of training signals.  Finally, we derive preferable system parameters for closed-loop training compared to open-loop training with memory.

\subsection{Open-loop training with memory}\label{lower_analysis}
In the proposed open-loop training with memory, we assume that the base station and the user share a common set of training signals that can be indexed with $B$ bits given by\footnote{Our framework can also be combined with a time-varying $\bP$ similar to how differential codebooks are used in CSI quantization \cite{tm_correlated4,tm_correlated5} for better performance.} $\cP=\left\{\bP_1,\ldots,\bP_{2^B}\right\}$.  Then, training signal for the $i$-th fading block $\bX_i$ is given as
\begin{equation}\label{rr_X}
  \bX_i = \bP_{\mathrm{mod}\left(i,2^B\right)+1},\quad i=0,\ldots,T-1,
\end{equation}
in a round-robin manner, which requires no feedback for the training signal from the user to the base station.  However, the user estimates the channel $\bh_i$ based not only on $\by_{i,train}$ but also on $\{\by_{k,train}\}_{k=0}^{i-1}$ and the channel statistics $\eta$ and $\bR$.  Note that this problem is similar to state prediction in dynamical systems.  With the training problem formulation, \eqref{h_model} specifies the state evolution and \eqref{eq_train} is the input-output equation \cite{Kay}.  Thus, the user can rely on the Kalman filter (or a more advanced filter such as the particle filter in \cite{particle1}) to track the channel evolution and provide a more accurate channel estimate.
\begin{table}
\caption{Sequential MMSE channel estimation based on the Kalman filter \cite{Kay}.}\label{kalman_table}
\hrule\hrule
\vspace{0.2cm}
\noindent \textbf{Initialization:}
\begin{align*}
 \widehat{\bh}_{0\mid -1}  &= \mathbf{0},\\
 \bR_{0\mid -1} &= \bR = E\left[\bh_0\bh_0^H\right].
\end{align*}

\noindent \textbf{Prediction:}
\begin{equation*}
  \widehat{\bh}_{i \mid i-1} = {\eta}\widehat{\bh}_{i-1 \mid i-1}.
\end{equation*}

\noindent \textbf{Minimum prediction MSE matrix ($N_t \times N_t$):}
\begin{alignat}{2}\notag
\bR_{i\mid i-1} = \eta^2 \bR_{i-1\mid i-1} + (1-\eta^2) \bR.
\end{alignat}

\noindent \textbf{Kalman gain matrix ($N_t \times T$):}
\begin{equation}\notag
\bK_{i} = \bR_{i\mid i-1} \bX_i\left(\bI_{T}+\bX_i^H\bR_{i\mid i-1}\bX_i\right)^{-1}.
\end{equation}

\noindent \textbf{Correction:}
\begin{equation}\notag
\widehat{\bh}_{i \mid i} =
\widehat{\bh}_{i \mid i-1} + \bK_i \left(\by_{i,train}-\bX_i^H\widehat{\bh}_{i \mid i-1} \right).
\end{equation}

\noindent \textbf{Minimum MSE matrix ($N_t \times N_t$):}
\begin{equation}\notag
\bR_{i\mid i} = \left(\bI_{N_t} - \bK_i\bX_i^H\right)\bR_{i\mid i-1}.
\end{equation}
\hrule\hrule
\end{table}

To begin with, we denote
\begin{equation*}
  \widehat{\bh}_{i_1\mid i_2} = E\left[\bh_{i_1}\mid \{\by_{k,train}\}_{k=0}^{i_2}\right]
\end{equation*}
as the predicted value of $\bh_{i_1}$ given $\{\by_{k,train}\}_{k=0}^{i_2}$ for $i_1\geq i_2$.  Then, we can define the sequential MMSE estimator $\widehat{\bh}_{i\mid i}$ based on $\{\by_{k,train}\}_{k=0}^{i}$ as in Table \ref{kalman_table}.  Note that the distribution of $\widehat{\bh}_{i\mid i}$ given $\{\by_{k,train}\}_{k=0}^{i-1}$ is complex Gaussian with mean $\widehat{\bh}_{i\mid i-1}$ and covariance
\begin{equation}\notag
\bR_{p,i} = \bR_{i\mid i-1} \bX_i \left(\bI_T + \bX_i^H\bR_{i \mid i-1} \bX_i \right)^{-1} \bX_i^H\bR_{i\mid i-1}.
\end{equation}
Because we assume perfect CSI feedback from the user to the base station, the beamforming vector becomes
\begin{equation}\label{w_mmse}
  \bw_i = \frac{\widehat{\bh}_{i \mid i}}{\|\widehat{\bh}_{i \mid i}\|}.
\end{equation}
From the numerical results in Section \ref{sim}, open-loop training with memory can significantly increase the channel estimation performance.

\subsection{Closed-loop training with memory}\label{cl_training}
\begin{figure*}[t]
\centering
\includegraphics[width=1.3\columnwidth]{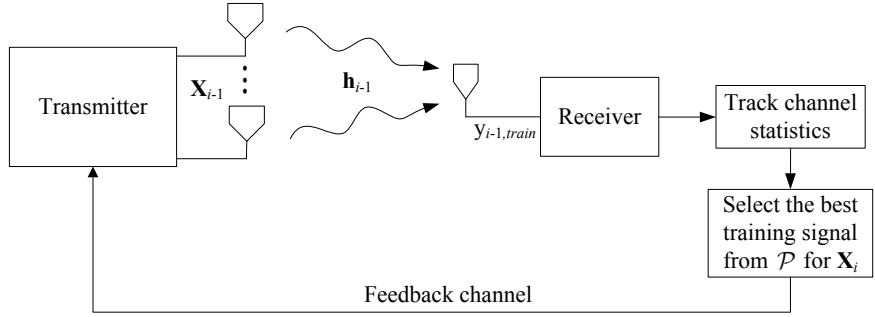}\\
\caption{Concept of closed-loop training.}\label{tracking}
\end{figure*}

We assume the channels are correlated in time and space.  Thus, the training signal at the $i$-th fading block can be adapted using the channel statistics and the previously received training signals $\{\by_{k,train}\}_{k=0}^{i-1}$ if they are available to the transmitter.  Because the base station will not have direct access to the channel statistics and $\{\by_{k,train}\}_{k=0}^{i-1}$, the best training signal $\bP_{i,\mathrm{best}}$ is selected from a predefined set of training signals $\cP=\left\{\bP_1,\ldots,\bP_{2^B}\right\}$ at the user and sent back to the base station with $B$ bits of feedback.  The base station then uses the fed back signal as the training signal for the $i$-th fading block.  The training signal selection at the user is based on using channel prediction to track the statistics of the channel at the $i$-th fading block conditioned on the user's side information as explained in open-loop training with memory.  The conceptual explanation of closed-loop training with memory is given in Fig. \ref{tracking}.

We propose two metrics for selecting $\bP_{i,\mathrm{best}}$ at the user, i.e., minimizing the MSE of channel estimation and maximizing the normalized average received SNR for the data communication phase.

\textit{1) Minimizing the MSE (MSE-based):} It is easy to show that the MSE between $\bh_i$ and $\widehat{\bh}_{i\mid i}$ is a function of $\bX_i$ and given as
\begin{align}
\nonumber \mathrm{MSE}\left(\bX_i\right) & =\frac{1}{N_t} E\left[\|\bh_i - \widehat{\bh}_{i\mid i}\|^2\right] \\
\nonumber & = \frac{1}{N_t}\mathrm{tr}\left(\bR_{i\mid i}\right)\\
& = \frac{1}{N_t}\mathrm{tr}\left(\bR_{i\mid i-1}-\bR_{p,i}\right).\label{tr_mse}
\end{align}
Therefore, the user selects $\bP_{i,\mathrm{best}}$ for the $i$-th block that minimizes the MSE as
\begin{align}
\bP_{i,\mathrm{best}} & = \argmin_{\bP_k\in\cP}\mathrm{MSE}\left(\bP_k\right)\label{mse_criterion}\\
\nonumber & = \argmax_{\bP_k\in\cP}\mathrm{tr}\left(\bR_{p,i}\right),%\label{tr_criterion}
\end{align}
and feeds back the $B$-bit index of $\bP_{i,\mathrm{best}}$ to the transmitter.  Then, the base station uses $\bX_i=\bP_{i,\mathrm{best}}$ for the training signal for the $i$-th block.
%
%Because we substitute $\bX_i$ with $\bP_k$ in \eqref{mse_criterion}, vectors and matrices defined in Table \ref{kalman_table} become functions of $\bP_k$ as well.  Same substitution will be applied throughout the paper, and it would be easy to understand from the context.

\textit{2) Maximizing the normalized average received SNR (SNR-based):} Using the beamforming vector $\bw = \frac{\widehat{\bh}_{i \mid i}}{\|\widehat{\bh}_{i \mid i}\|}$ as in \eqref{w_mmse}, $\Gamma_i$ given $\{\bX_k,\by_{k,train}\}_{k=0}^{i}$ becomes
\begin{align}
\nonumber \Gamma_i\left(\{\bX_k,\by_{k,train}\}_{k=0}^{i}\right) &= E\left[\left|\bh_i^H\bw\right|^2\mid \{\bX_k,\by_{k,train}\}_{k=0}^{i}\right] \\
\nonumber  &= \bw^H\left(\widehat{\bh}_{i\mid i}\widehat{\bh}_{i\mid i}^H+\bR_{i\mid i}\right)\bw \\
 &= \left\|\widehat\bh_{i\mid i}\right\|^2 + \frac{\widehat\bh_{i\mid i}^H\bR_{i\mid i}\widehat\bh_{i\mid i}}{\left\|\widehat\bh_{i\mid i}\right\|^2}\label{rr_snr}.
\end{align}
The user maximizes the expected value of \eqref{rr_snr} averaged over $\by_{i,train}$ by selecting $\bP_{i,\mathrm{best}}$ as
\begin{align}
\nonumber  \bP_{i,\mathrm{best}}&=\argmax_{\bP_i\in\cP}E\left[\Gamma_i\left(\bP_i,\by_{i,train}\right)|\{\bX_k,\by_{k,train}\}_{k=0}^{i-1}\right]\\
&=\argmax_{\bP_i\in\cP}\Gamma_i\left(\bP_i,\{\bX_k,\by_{k,train}\}_{k=0}^{i-1}\right),\label{snr_expect}
\end{align}
with the expectation taken over $\by_{i,train}$.

We can evaluate $\Gamma_i\left(\bP_i,\{\bX_k,\by_{k,train}\}_{k=0}^{i-1}\right)$ in \eqref{snr_expect} as
\begin{equation*}
  \Gamma_i\left(\bP_i,\{\bX_k,\by_{k,train}\}_{k=0}^{i-1}\right)=\trace\left(\bR_{p,i}\right)+ \left\|\widehat{\bh}_{i\mid i-1}\right\|^2+q(\bP_i)
\end{equation*}
where $q(\bP_i)$ is defined as
\begin{equation*}
  q(\bP_i) = E\left[\left.\frac{\widehat\bh_{i\mid i}^H\bR_{i\mid i}\widehat\bh_{i\mid i}}{\left\|\widehat\bh_{i\mid i}\right\|^2}\right| \{\bX_k,\by_{k,train}\}_{k=0}^{i-1}\right].
\end{equation*}
 By defining $\alpha_1 = \widehat\bh_{i\mid i}^H\bR_{i\mid i}\widehat\bh_{i\mid i}$ and $\alpha_2 = \left\|\widehat\bh_{i\mid i}\right\|^2$, we can approximate $q(\bP_i)$ as \cite{Paolella}
\begin{equation}\notag
q(\bP_i) \approx \frac{E\left[\alpha_1\right]}{E\left[\alpha_2\right]}\left(1-\frac{Cov(\alpha_1,\alpha_2)}{E\left[\alpha_1\right]\cdot E\left[\alpha_2\right]}+\frac{Var(\alpha_2)}{\left(E\left[\alpha_2\right]\right)^2}\right)
\end{equation}
where
\begin{align*}
E\left[\alpha_1\right]&=\widehat\bh_{i\mid i-1}^H\bR_{i\mid
i}\widehat\bh_{i\mid i-1} +  \trace\left(\bR_{i\mid i}\bR_{p,i}\right),\\
E\left[\alpha_2\right]&=\|\widehat\bh_{i\mid i-1}\|^2 + \trace\left(\bR_{p,i}\right),\\
Var(\alpha_2) &= 4\widehat\bh_{i\mid i-1}^H \bR_{p,i}\widehat\bh_{i\mid
i-1} + 2\trace\left(\bR_{p,i}\right)^2,\\
Cov(\alpha_1,\alpha_2) &= 4\widehat\bh_{i\mid i-1}^H \bR_{i\mid
i}\bR_{p,i}\widehat\bh_{i\mid i-1} + 2\trace\left(\bR_{i\mid
i}\bR_{p,i}\bR_{p,i}\right).
\end{align*}
Thus, $\bP_{i,\mathrm{best}}$ can be selected by the user according to
\begin{equation}\label{optfunction2}
\bP_{i,\mathrm{best}} = \argmax_{\bP_i\in\cP}\left(\trace{\left(\bR_{p,i}\right)} +\left\|\widehat{\bh}_{i\mid i-1}\right\|^2+ q(\bP_i)\right),
\end{equation}
and the $B$-bit index of $\bP_{i,\mathrm{best}}$ can be sent as feedback from the user to the base station.

Note that maximizing \eqref{optfunction2} is the same as minimizing the MSE in \eqref{mse_criterion} augmented with the term $q(\bP_i)$ ($\widehat{\bh}_{i\mid i-1}$ is a constant regardless of $\bP_i$).  Numerical studies in Section \ref{sim} show that $q(\bP_i)$ has a non-negligible impact on the received SNR when $N_t$ is moderately large, the channel is highly correlated in space, and the SNR is low.  For other cases, however, the difference between the two metrics is negligible.

\subsection{Closed-loop training with memory with full feedback to minimize MSE}\label{upper_analysis}
In this subsection, we derive the optimal training signal $\bX_{i,\mathrm{opt}}$ of closed-loop training with memory that minimizes the MSE of the $i$-th fading block in \eqref{tr_mse}.  Note that $\bX_{i,\mathrm{opt}}$ is possible only when closed-loop training supports unlimited feedback overhead.  Thus, $\bX_{i,\mathrm{opt}}$ only gives an MSE lower bound of the proposed closed-loop with memory.

Because the MSE in \eqref{tr_mse} has the same formulation as \eqref{mse_orig} once $\bR_{i\mid i-1}$ is replaced by $\bR$, the same arguments employed in Lemma \ref{opt_X} can be used to show that the optimal training signal is given as
\begin{equation}\label{Xopt_kalman}
  \bX_{i,\mathrm{opt}} = \sqrt{\rho}\bU_{i[1:T]}
\end{equation}
where $\bR_{i\mid i-1}=\bU_i\mathbf{\Lambda}_i\bU_i^H$ with $\mathbf{\Lambda}_i=\diag\left(\left[\lambda_{i,1},\cdots,\lambda_{i,N_t}\right]\right)$.  Comparing \eqref{Xopt_orig} and \eqref{Xopt_kalman}, the optimal training signal is now the first $T$ dominant eigen-directions of the prediction matrix $\bR_{i\mid i-1}$.

It is interesting to point out that, using the recursive derivation of $\bR_{i\mid i-1}$ and $\bR_{i\mid i}$, we can easily show that $\bU_i$ is column-wise permutation of $\bU$ (the eigenvector matrix of $\bR$), which means the $T$ dominant eigenvectors varies with $i$.  Thus, the full-feedback scheme can be thought of as a training technique that scans among the eigen-directions of the original spatial correlation matrix $\bR$.  This property has been exploited in \cite{song} for FDD massive MIMO training when the base station has perfect knowledge of $\bR$.

We now derive the MSE of the $i$-th fading block using $\bX_{i,\mathrm{opt}}$ to provide a lower bound on the MSE of closed-loop training with memory in the following lemma.
\begin{lemma}\label{mse_lower}
Recall $\bR=\bU\mathbf{\Lambda}\bU^H$ and let $\bU_0=\bU$ and $\mathbf{\Lambda}_0=\mathbf{\Lambda}$.  Using the Kalman filter update in Table \ref{kalman_table} and the optimal training signal $\bX_{i,\mathrm{opt}} = \sqrt{\rho}\bU_{i[1:T]}$, the MSE at the $i$-th fading block is given as
\begin{equation}\label{mse_lower_eq}
  \mathrm{MSE}_i(\bX_{i,\mathrm{opt}}) = 1-\frac{1}{N_t} \sum_{k=0}^{i}\sum_{t=1}^{T}\frac{\eta^{2(i-k)}\rho \lambda_{k,t}^2}{\rho \lambda_{k,t}+1},
\end{equation}
where $\lambda_{k,t}$ is the $t$-th dominant eigenvalue of $\bR_{k|k-1}$.
\end{lemma}
\begin{IEEEproof}
See Appendix D.
\end{IEEEproof}
When $i=0$, \eqref{mse_lower_eq} simplifies down to \eqref{tr_calc}.  Lemma \ref{mse_lower} clearly shows that in temporally correlated channels with $\eta \approx 1$, the $\mathrm{MSE}_i$ in \eqref{mse_lower_eq} is always lower than the MSE of closed-loop/single-shot training in \eqref{tr_calc} for $i>0$.  Thus, channel prediction with an optimized training signal selection will improve channel estimation performance.  Although it is hard to analyze the normalized received SNR with $\bX_{i,\mathrm{opt}}$, we can expect from Lemma \ref{mse_lower} that closed-loop training with memory can effectively reduce the ceiling effect of single-shot training discussed in Section \ref{ceiling_effet}.

\subsection{Design of training signal set $\cP$}
Now, we discuss an effective way of generating a set of training signals $\cP$.  We again restrict $\cP$ to be a subset of the set $\cX$ meaning
\begin{equation*}
  \cP \subset \cX = \left\{\bF:\bF\in \mathbb{C}^{N_t \times T},~\bF^H\bF = \rho \bI_T\right\}.
\end{equation*}
It is shown in the previous subsection that the optimal training method that minimizes the MSE scans over the eigen-directions of $\bR$ that are orthogonal to each other.  To mimic this, the training signals in $\cP$ should be as orthogonal as possible.  This can be numerically achieved by Grassmannian subspace packing (GSP).

The chordal distance between the two matrices $\bX$ and $\bY$ is given as
\begin{equation*}
  d_c\left(\bX,\bY\right)\triangleq \frac{1}{\sqrt{2}}\left\|\bX\bX^H-\bY\bY^H\right\|_F,
\end{equation*}
and the minimum chordal distance of a candidate training set $\cP_c=\left\{\bP_{c,1},\dots,\bP_{c,2^B}\right\}$ as
\begin{equation*}
  d_{c,\min}\left(\cP_c\right)\triangleq \min_{1\leq m\leq n \leq 2^B}d_c\left(\bP_{c,m},\bP_{c,n}\right).
\end{equation*}
Then, the GSP training set $\cP_{\mathrm{GSP}}$ can be given as
\begin{equation*}
  \cP_{\mathrm{GSP}} = \argmax_{\cP_c\subset\cP_{\mathrm{all}}}d_{c,\min}\left(\cP_c\right),
\end{equation*}
where $\cP_{\mathrm{all}}$ is a set of all possible candidate sets $\cP_c$.  We adopt numerically optimized $\cP_{\mathrm{GSP}}$ for performance evaluation in Section \ref{sim}.

\subsection{Impact of system parameters on closed-loop training}\label{impact_cl}
In this subsection, we give explanations of scenarios when closed-loop training with memory has a gain compared to open-loop training with memory.  The explanations are based on the optimal training signal $\bX_{i,\mathrm{opt}}$ that minimizes the MSE for tractable analyses.

\textit{1) Variation with SNR ($\rho$):} The minimization of the MSE in \eqref{tr_mse} can be first converted to the maximization of
\begin{equation*}
\mathrm{tr}\left(\bR_{i\mid i-1} \bX_i \left(\bI_T + \bX_i^H\bR_{i \mid i-1} \bX_i \right)^{-1} \bX_i^H\bR_{i\mid i-1}\right)
\end{equation*}
and approximated as
\begin{equation*}
  \mathrm{tr}\left(\bX_i^H\bR_{i\mid i-1}^2\bX_i\right)
\end{equation*}
in the low-SNR regime and
\begin{equation*}
    \mathrm{tr}\left(\left(\bX_i^H\bR_{i\mid i-1}\bX_i\right)^{-1}\bX_i^H\bR_{i\mid i-1}^2\bX_i\right)
\end{equation*}
in the high-SNR regime.  The optimal training signal in both cases is again $\bX_{i,\mathrm{opt}} = \sqrt{\rho} \bU_{i[1:T]}$.  However, if we plug in $\bX_{i,\mathrm{opt}}$ into each approximated objective function, we have $\sum\limits_{t=1}^{T}\rho \lambda_{i,t}^2$ in the low-SNR regime and $\sum\limits_{t=1}^{T}\rho \lambda_{i,t}$ in the high-SNR regime.  Assuming $\lambda_{i,t} > 1$ for $t=1,\ldots,T$, which is typically true for spatially correlated massive MIMO channels, the subspace spanned by the columns of the training signal is more important in the low-SNR regime than the high-SNR regime.  Thus, it is expected that closed-loop training with memory would be more beneficial in the low-SNR regime.

\textit{2) Variation with length of training phase ($T$):} When $T=1$, the direction of the optimal training signal is the dominant eigenvector of $\bR_{i\mid i-1}$ so that $\bx_{i,\mathrm{opt}}=\bU_{i[1:1]}$.  However, when $T=N_t$, it is easy to show in a similar manner as \eqref{tr_calc} that any scaled unitary matrix $\bX_{i,\mathrm{opt}}=\sqrt{\rho}\bV$ is optimal giving
\begin{align*}
  &\mathrm{tr}\left(\left(\bI_{N_t} + \bX_{i,\mathrm{opt}}^H\bR_{i \mid i-1} \bX_{i,\mathrm{opt}} \right)^{-1} \bX_{i,\mathrm{opt}}^H\bR_{i\mid i-1}^2\bX_{i,\mathrm{opt}}\right)\\
  &\quad=\sum_{t=1}^{N_t}\frac{\rho \lambda_{i,t}^2}{\rho \lambda_{i,t} +1}.
\end{align*}
This means that there is no preferable direction for $\bX_{i,\mathrm{opt}}$ when $T=N_t$.  In the case of $1<T<N_t$, it is obvious that any other combination of $T$ columns of $\bU_i$ (except rearranging the first $T$ columns of $\bU_i$) for $\bX_i$ gives inferior results than $\bU_{i[1:T]}$.  However, the gap between $\bU_{i[1:T]}$ and other combinations will reduce as $T$ increases.  Thus, the subspace spanned by the columns of the training signal is more important when $T$ (or the ratio $\frac{T}{N_t}$) is small, and closed-loop training with memory is most beneficial in this scenario.

\textit{3) Variation with fading block index $i$:} Intuitively, the subspace spanned by the columns of the training signal seems to be more important at the beginning of channel estimation when the user lacks accurate channel knowledge.  To explain this rigorously, we know from \eqref{mse_lower_eq} in Lemma \ref{mse_lower} that the MSE is decreased by $\sum\limits_{t=1}^{T}\frac{\rho\lambda^2_{i,t}}{\rho\lambda_{i,t}+1}$ at the $i$-th block when $\bX_{i,\mathrm{opt}}$ is used as a training signal.  We also know from the proof of Lemma \ref{mse_lower} that the first $T$ eigenvalues of $\bR_{i\mid i-1}$ are decreasing with $i$ such that $\lambda_{i-1,t}\geq \lambda_{i,t}$ for $t=1,\ldots,T$. Because of the Schur-convexity of $\sum\limits_{t=1}^{T}\frac{\rho\lambda^2_{i,t}}{\rho\lambda_{i,t}+1}$ as shown in Appendix B, we have
\begin{equation*}
  \sum_{t=1}^{T}\frac{\rho\lambda^2_{i-1,t}}{\rho\lambda_{i-1,t}+1}\geq \sum_{t=1}^{T}\frac{\rho\lambda^2_{i,t}}{\rho\lambda_{i,t}+1}.
\end{equation*}
Thus, having the right subspace spanned by the columns of the training signal can reduce the MSE more effectively when $i$ is small, and closed-loop training with memory has more gain when a prior channel estimate is not accurate at the beginning of channel estimation.

\section{Simulation Results}\label{sim}
\begin{figure}[t]
\centering
\includegraphics[width=0.9\columnwidth]{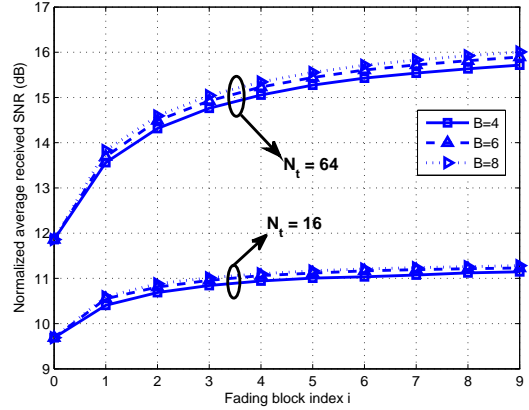}\\
\caption{$\Gamma_i^{(\mathrm{dB})}$ of SNR-based closed-loop training according to the fading block index $i$ with $\rho=0$dB, $T=2$, $a=0.9$, and different $B$ and $N_t$ values.}\label{snr_diff_B}
\end{figure}

\begin{figure*}
\centering
\subfloat[$\rho=0$dB, $T=1$.]{
\includegraphics[width=0.9\columnwidth]{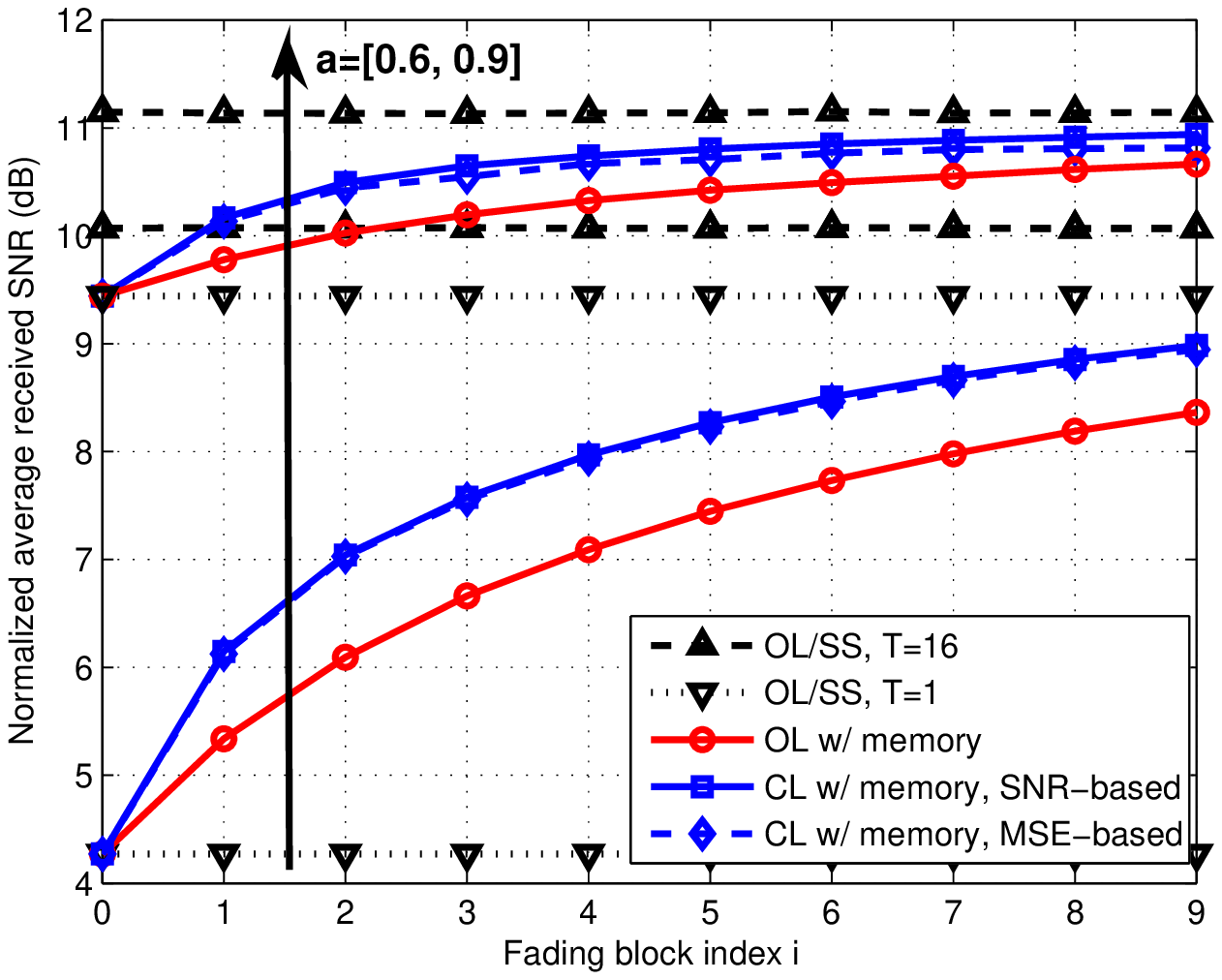}
\label{16Tx0dB1T_snr}
}
\subfloat[$\rho=0$dB, $T=2$.]{
\includegraphics[width=0.9\columnwidth]{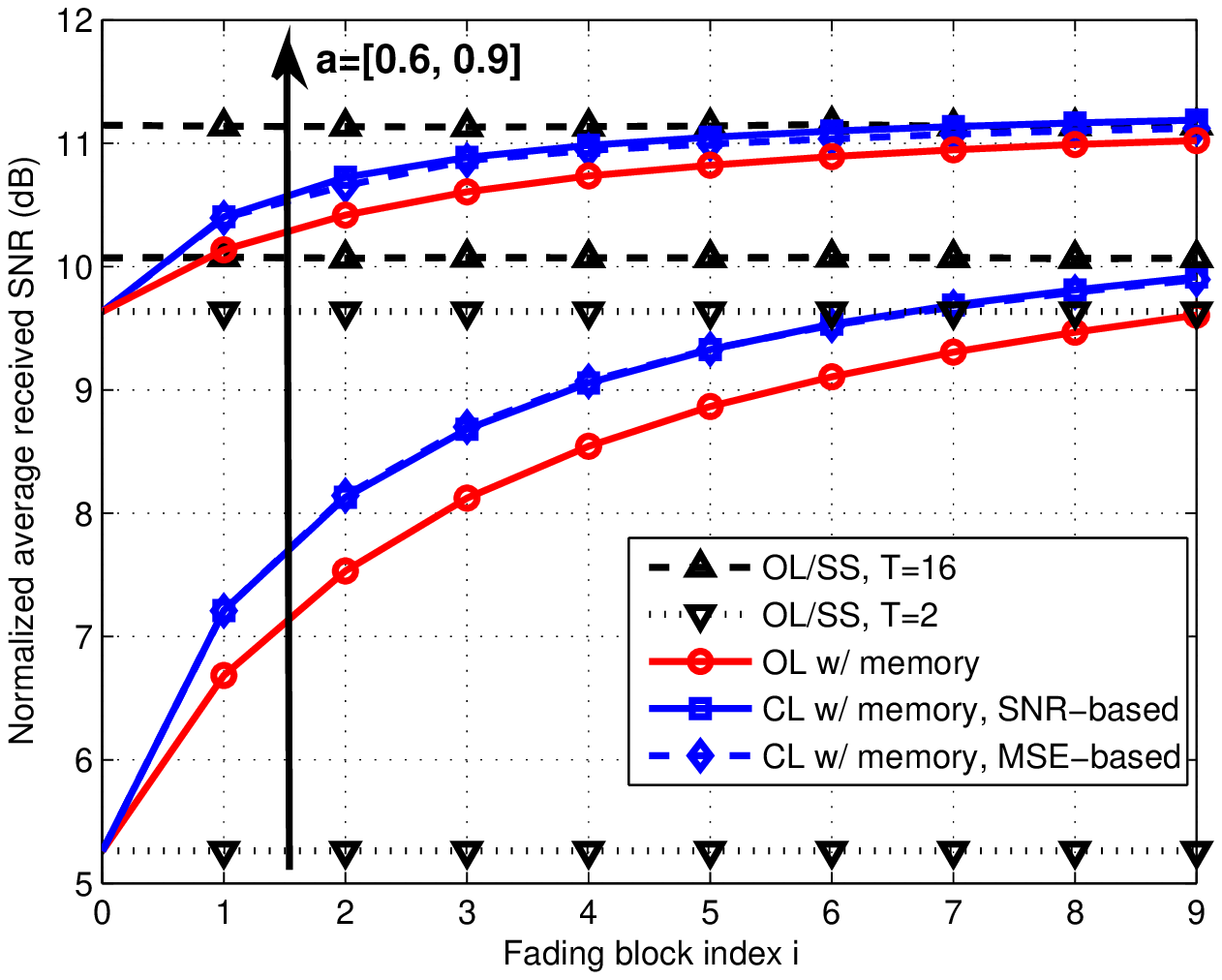}
\label{16Tx0dB2T_snr}
}\\
\subfloat[$\rho=20$dB, $T=1$.]{
\includegraphics[width=0.9\columnwidth]{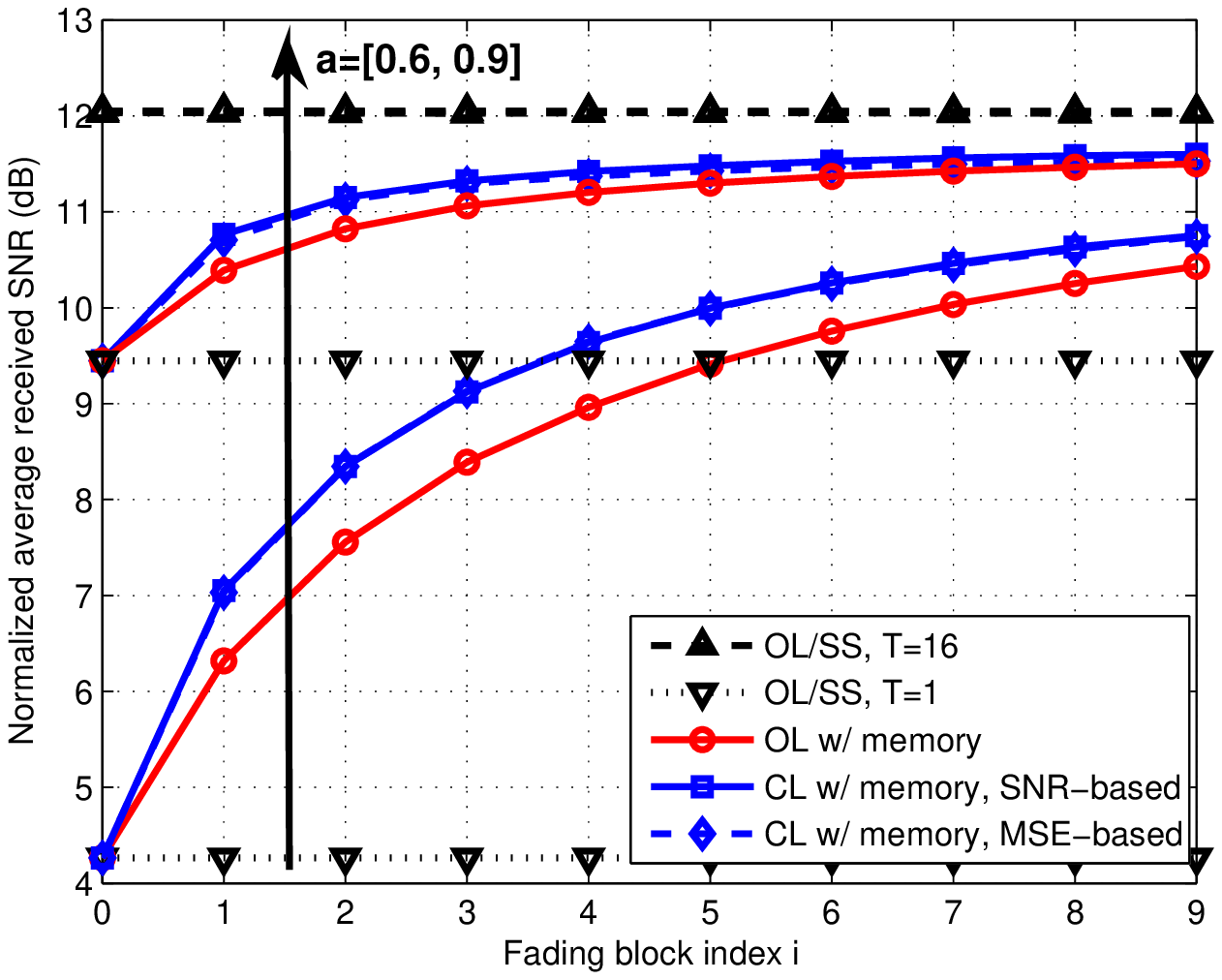}
\label{16Tx20dB1T_snr}
}
\subfloat[$\rho=20$dB, $T=2$.]{
\includegraphics[width=0.9\columnwidth]{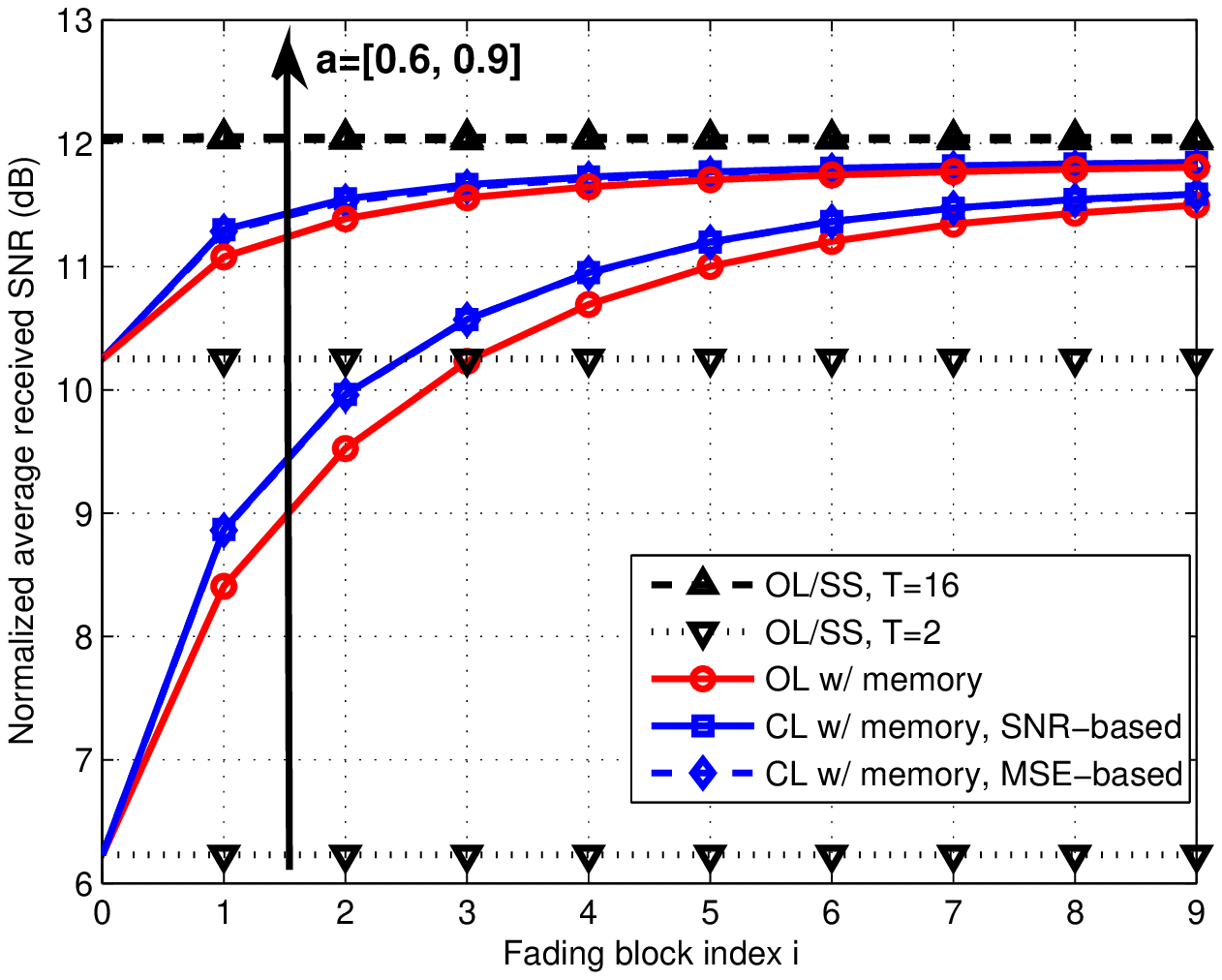}
\label{16Tx20dB2T_snr}
}
\caption{$\Gamma_i^{(\mathrm{dB})}$ according to the fading block index $i$ with $N_t=16$ and different $\rho$, $T$, and $a$ values.}
\label{16Tx_snr}
\end{figure*}

To evaluate the proposed training frameworks, we present Monte-Carlo simulation results with 10000 iterations in this section.  Each iteration consists of 10 fading blocks which are temporally and spatially correlated as shown in \eqref{h_model}.  We adopt Jakes' model \cite{Prok} for the temporal correlation coefficient $\eta = J_0(2\pi f_D \tau)$ where $J_0(\cdot)$ is the $0$-th order Bessel function of the first kind, $\tau=5\mathrm{ms}$ is the channel instantiation interval, and $f_D=\frac{vf_c}{c}$ denotes the maximum Doppler frequency.  With the user speed $v=3\mathrm{km/h}$, the carrier frequency $f_c = 2.5\mathrm{GHz}$, and the speed of light $c=3\times 10^8\mathrm{m/s}$, the temporal correlation coefficient becomes $\eta=0.9881$.  Assuming a 5ms coherence time and the frame structure of 3GPP LTE FDD systems \cite{lte}, each fading block consists of $L\approx 10$ static channel uses.  We adopt the same spatial correlation matrix $\bR$ as in \eqref{exp_model}, and the numerically optimized GSP training set $\cP_{\mathrm{GSP}}$ that is used in both open-loop and closed-loop training with memory.  The dB scale of the normalized average received SNR in \eqref{snr_average}, $\Gamma_i^{(\mathrm{dB})}$, is used for the performance metric.

We first compare $\Gamma_i^{(\mathrm{dB})}$ of closed-loop training with memory based on the SNR metric with different values of $B$ for $\cP_{\mathrm{GSP}}$ in Fig. \ref{snr_diff_B}.  We set the signal power $\rho=0$dB, the number of channel uses for training $T=2$, and the spatial correlation parameter $a=0.9$.  As the size of $\cP_{\mathrm{GSP}}$ increases, $\Gamma_i^{(\mathrm{dB})}$ also increases in both $N_t=16$ and 64 cases.  The gain of having larger $B$ is more prominent when $N_t$ is large; however, it is expected that having $B$ less than 10 bits seems to be enough to have a notable gain.  This means that the computational complexity of training signal selection might not be a big issue in practice.  We set $B=6$ for other simulations in this section.

\begin{figure*}
\centering
\subfloat[$\rho=0$dB, $T=2$.]{
\includegraphics[width=0.9\columnwidth]{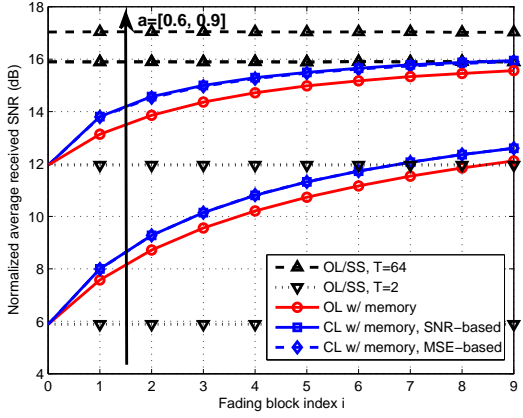}
\label{64Tx0dB2T_snr}
}
\subfloat[$\rho=0$dB, $T=4$.]{
\includegraphics[width=0.9\columnwidth]{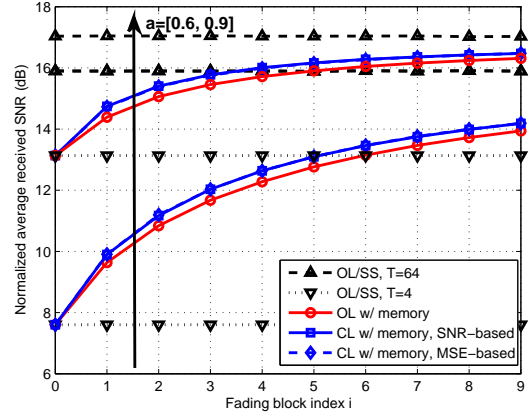}
\label{64Tx0dB4T_snr}
}\\
\subfloat[$\rho=20$dB, $T=2$.]{
\includegraphics[width=0.9\columnwidth]{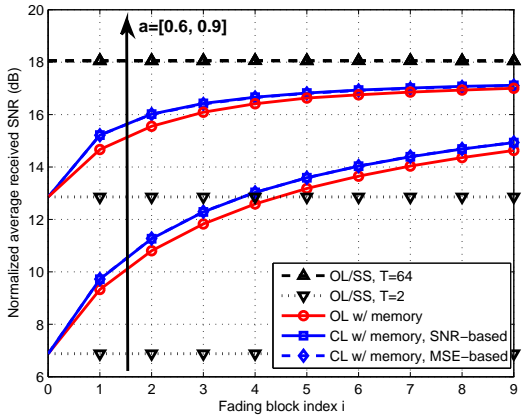}
\label{64Tx20dB2T_snr}
}
\subfloat[$\rho=20$dB, $T=4$.]{
\includegraphics[width=0.9\columnwidth]{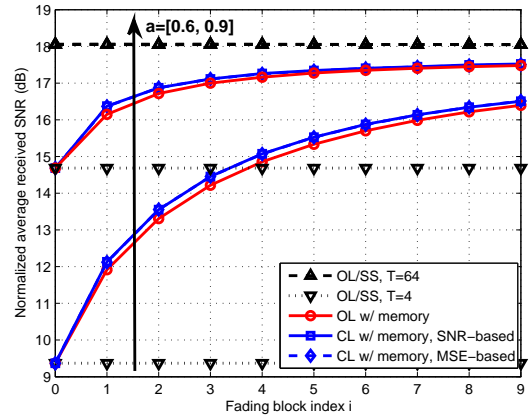}
\label{64Tx20dB4T_snr}
}
\caption{$\Gamma_i^{(\mathrm{dB})}$ according to the fading block index $i$ with $N_t=64$ and different $\rho$, $T$, and $a$ values.}
\label{64Tx_snr}
\end{figure*}

In Figs. \ref{16Tx_snr} and \ref{64Tx_snr}, we plot $\Gamma_i^{(\mathrm{dB})}$ of
open-loop/single-shot (OL/SS), open-loop with memory (OL w/ memory), and closed-loop with memory based on the MSE metric (CL w/ memory, MSE-based) and the SNR metric (CL w/ memory, SNR-based) training schemes according to the fading block index $i$ with $N_t=16$ and $64$ and different values of $\rho$, $T$, and $a$.  We randomly reorder the indices of $\cP_{\mathrm{GSP}}$ at each iteration to preclude the effect of a specific ordering of training signals in open-loop training.

From the figures, it is easy to verify that with the same $T$ the proposed training frameworks outperform open-loop/single-shot training, which adopts the training signal for the $i$-th block with a round-robin manner in \eqref{rr_X} with $\cP_{\mathrm{GSP}}$ but only relies on $\by_{i,train}$ for the $i$-th block channel estimation.  Moreover, in Fig. \ref{16Tx0dB2T_snr}, closed-loop training with memory is slightly better than open-loop/single-shot training with $T=N_t$ when $a=0.9$ and $i=9$.  This shows that the successive channel estimation approach of closed-loop training with memory can effectively alleviate the impact of noise.
%two column
\begin{figure*}
\centering
\subfloat[$N_t=16$, $\rho=0$dB, $T=2$.]{
\includegraphics[width=0.9\columnwidth]{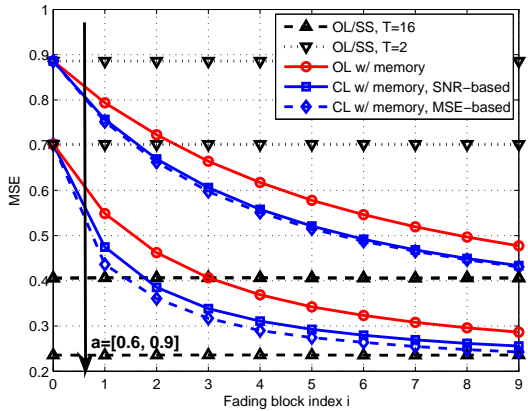}
}
\subfloat[$N_t=64$, $\rho=20$dB, $T=4$.]{
\includegraphics[width=0.9\columnwidth]{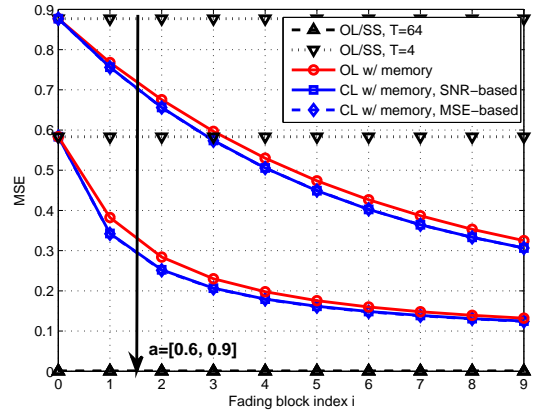}
}
\caption{MSE according to the fading block index $i$ with different $N_t$, $\rho$, $T$, and $a$ values.}
\label{mse_simul}
\end{figure*}

Comparing open-loop and closed-loop training with memory, the gain of closed-loop training with memory becomes larger when 1) $\rho$ is low, 2) $T$ is small relative to $N_t$, and 3) $i$ is small, which are inline with the discussions in Section \ref{impact_cl}.  It is shown in Figs. \ref{16Tx_snr} that closed-loop training with memory based on the SNR metric gives non-negligible gain compared to closed-loop training based on the MSE metric when $N_t$ is moderately large and $\rho$ is small in highly correlated case.

The performance of all schemes increases as $a$ increases, i.e., when channels are highly correlated in space.  This certainly shows that the spatial correlation \textit{helps} in estimating the channel, which is pointed out in Lemma \ref{sp_help} and \cite{training2,training5}.
%two column
\begin{figure*}
\centering
\subfloat[$\rho=0$dB, $T=2$.]{
\includegraphics[width=0.9\columnwidth]{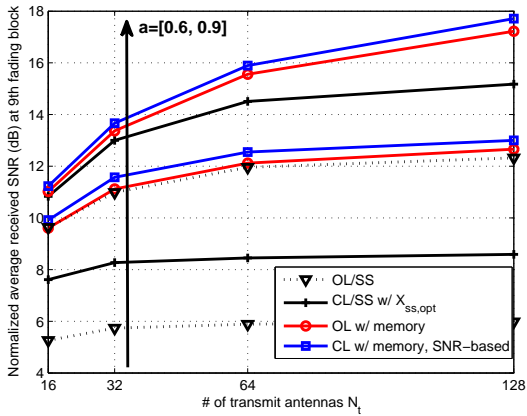}
}
\subfloat[$\rho=20$dB, $T=4$.]{
\includegraphics[width=0.9\columnwidth]{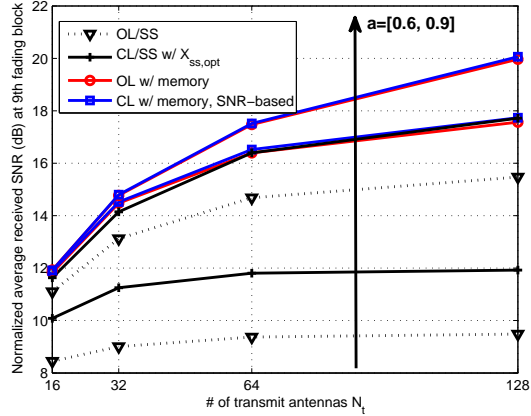}
}
\caption{$\Gamma_i^{(\mathrm{dB})}$ according to $N_t$ with different $\rho$, $T$, and $a$ values.}
\label{tx_ant}
\end{figure*}

We also plot the MSE of each scheme in Fig. \ref{mse_simul}.  Similar to the previous figures of $\Gamma_i^{(\mathrm{dB})}$, the proposed training frameworks give far lower MSE than open-loop/single-shot training.  Note that the MSE of closed-loop training with memory based on the MSE metric is smaller than that of closed-loop training based on the SNR metric when $N_t=16$, which shows the tradeoff between SNR and MSE metric in closed-loop training.

In Fig. \ref{tx_ant}, we plot $\Gamma_i^{(\mathrm{dB})}$ of the 9th fading block according to $N_t$.  Note that $\Gamma_i^{(\mathrm{dB})}$ of open-loop/single-shot training quickly saturates as $N_t$ increases. We also plot the results of closed-loop/single-shot training with full feedback of $\bX_{\mathrm{ss,opt}}$ (CL/SS w/$\bX_{\mathrm{ss,opt}}$) discussed in Section \ref{sat_effect}, which also experiences the ceiling effect, for comparison.  It is obvious that open-loop and closed-loop training with memory can effectively reduce the ceiling effect even with small $T$ compared to $L$, especially when $a$ is large.  This clearly shows that the gain of the proposed training schemes for massive MIMO systems.

\begin{figure}[t]
\centering
\includegraphics[width=0.9\columnwidth]{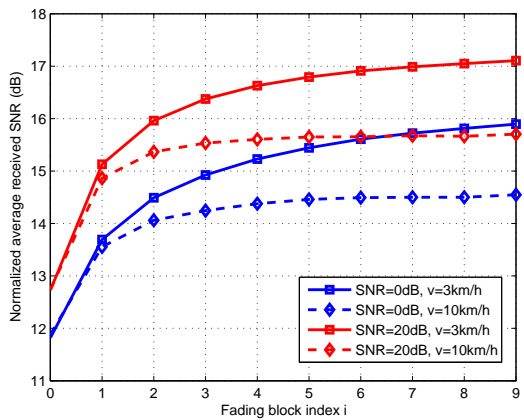}\\
\caption{$\Gamma_i^{(\mathrm{dB})}$ of SNR-based closed-loop training according to the fading block index $i$ with $T=2$, $a=0.9$, $B=6$, $N_t=64$ and different $\rho$ and $v$ values.}\label{snr_diff_vel}
\end{figure}
Finally, we plot closed-loop training with memory based on the SNR metric with different user velocities in Fig. \ref{snr_diff_vel}.  Note that $v=10\mathrm{km/h}$ corresponds to $\eta=0.8721$.  The loss from the high velocity is severe, i.e., almost 1.4dB loss of the received SNR in the saturation regime.  When the user velocity is high, instead of relying on the closed-loop training framework, the base station should transmit sounding signals more frequently in an open-loop manner in practice.  For example, four sounding signals (or reference signals) would be transmitted within a 1ms time period to support 350km/h user velocity in 3GPP LTE systems \cite{lte}.  Even in this case, the proposed open-loop training with memory can be exploited.

\section{Conclusion}\label{conclusion}
In this paper, we proposed open and closed-loop training frameworks using successive channel prediction/estimation at the user for FDD massive MIMO systems.  By exploiting prior channel information such as the long-term channel statistics and previous received training signals at the user, channel estimation performance can be significantly improved with only small length of training signals in each fading block compared to open-loop/single-shot training.  Moreover, with a small amount of feedback, which indicates the best training signal to be sent for the next fading block, from the user to the base station, the downlink training overhead can be further reduced even when the transmitter lacks any kind of side information, e.g., statistics of the channel.

\appendices
\section{Proof of Lemma 1}
Because $\bR$ is fixed, minimizing the MSE problem can be converted to
\begin{align}
\nonumber &\argmin_{\bX\in\cX}\mathrm{MSE}\left(\bX\right)\\
\nonumber &\quad = \argmax_{\bX\in\cX}\mathrm{tr}\left(\bR \bX \left(\bI_T + \bX^H\bR \bX\right)^{-1} \bX^H\bR\right)\\
&\quad\stackrel{(a)}{=} \argmax_{\bX\in\cX}\mathrm{tr}\left(\left(\bI_T + \bX^H\bR \bX \right)^{-1} \bX^H\bR^2\bX\right),\label{mse_optimization}
\end{align}
where $(a)$ is from the fact that $\trace(\bA \bB \bC)=\trace(\bB \bC \bA )$.  Using the eigen-decomposition of $\bR=\bU \mathbf{\Lambda} \bU^H$, we can rewrite \eqref{mse_optimization} as
\begin{align*}
&\argmin_{\bX\in\cX}\mathrm{MSE}\left(\bX\right)\\
&~=\argmax_{\bX\in\cX}\mathrm{tr}\left(\left(\bI_T + \bX^H\bU \mathbf{\Lambda} \bU^H \bX \right)^{-1} \bX^H \bU \mathbf{\Lambda}^2 \bU^H \bX\right)\\
&~\stackrel{(a)}{=} \argmax_{\widetilde{\bX}\in\cX}\mathrm{tr}\left(\left(\bI_T +\widetilde{\bX}^H \mathbf{\Lambda} \widetilde{\bX} \right)^{-1} \widetilde{\bX}^H \mathbf{\Lambda}^2 \widetilde{\bX}\right) \\
  &~\stackrel{(b)}{=} \argmax_{\widetilde{\bX}\in\cX}\mathrm{tr}\left(\left(\widetilde{\bX}^H \left(\frac{1}{\rho}\bI_{N_t} + \mathbf{\Lambda} \right)\widetilde{\bX} \right)^{-1} \widetilde{\bX}^H \mathbf{\Lambda}^2 \widetilde{\bX}\right)
\end{align*}
where $(a)$ comes from the change of the variable $\widetilde{\bX}=\bU^H \bX$, and $(b)$ is from $\widetilde{\bX}^H \widetilde{\bX}= \rho \bI_T$.  Because $\rho^{-1}\bI_{N_t}+\mathbf{\Lambda}$ and $\mathbf{\Lambda}^2$ are all real diagonal matrices with strictly positive entries in decreasing order, from the property of the block generalized Rayleigh quotient \cite{baker_phd}, the optimal solution for single-shot training is given as
\begin{equation*}
  \widetilde{\bX}_{\mathrm{ss,opt}} = \sqrt{\rho}\bI_{N_t[1:T]}.
\end{equation*}
Thus, $\bX_{\mathrm{ss,opt}}$ becomes
\begin{equation*}
  \bX_{\mathrm{ss,opt}} = \bU \widetilde{\bX}_{\mathrm{ss,opt}} = \sqrt{\rho} \bU \bI_{N_t[1:T]} = \sqrt{\rho} \bU_{[1:T]},
\end{equation*}
which finishes the proof.
\qed

\section{Proof of Lemma 2}
The basic concept of majorization theory which is used to prove Lemma \ref{sp_help} is from \cite{training2,majo}.

Let the real-valued function $f~:~\mathbb{R}^{T} \rightarrow \mathbb{R}$ as
\begin{equation*}
  f(\bx) = \sum_{t=1}^{T}\frac{\rho x_t^2}{\rho x_t+1}
\end{equation*}
with a vector $\bx=\left[x_{1},x_{2},\cdots,x_{T}\right]^T$ and a constant $\rho>0$.  Note that $f(\bx)$ is the same as the second term in \eqref{tr_calc}, which should be maximized to minimize the MSE.  It is easy to show that $f(\bx)$ is Schur-convex because $f(\bx)$ is symmetric and $\frac{\rho x_t^2}{\rho x_t+1}$ is convex.  By majorization theory and the property of Schur-convexity, we have
\begin{equation*}
  \bx\succ \by \Rightarrow f(\bx)\geq f(\by)
\end{equation*}
with arbitrary two vectors $\bx,\by\in \mathbb{R}^T$.  Because we assume $\lambda\left(\bR_H\right)\succ \lambda\left(\bR_L\right)$, we have $\mathrm{MSE}\left(\bX_H\right) \leq \mathrm{MSE}\left(\bX_L\right)$.
\qed

\section{Proof of Lemma 3}
First, we decompose $\bh=\widehat{\bh}+\br$ where $\widehat{\bh}$ and $\br$ are independent because of the orthogonality of the MMSE estimator \cite{Kay}.  Note that the covariance of $\br$ is given as
\begin{equation*}
\bR_{\br} = \bR-\bR\bX \left(\bI_T + \bX^H\bR \bX \right)^{-1} \bX^H\bR.
\end{equation*}
We let $\bR_{\br}=\bU_{\br}\mathbf{\Lambda}_{\br}\bU_{\br}$ where $\bU_{\br}$ is the eigenvector matrix and $\mathbf{\Lambda}_{\br}=\diag\left(\left[\lambda_{\br,1},\cdots,\lambda_{\br,N_t}\right]\right)$ is the eigenvalue matrices of $\bR_{\br}$ in decreasing order.  Now, we expand $\Gamma_{\mathrm{ss,opt}}$ as
\begin{align*}
\Gamma_{\mathrm{ss,opt}} &\stackrel{(a)}{=} E\left[E\left[\left|\bh \bw\right|^2 | \bh\right]\right]\\
& = E\left[\bw^H\left(\widehat{\bh}\widehat{\bh}^H+\bR_{\br}\right)\bw\right] \\
& \stackrel{(b)}{=} \trace\left(\bR_{\widehat{\bh}}\right)+E\left[\frac{\widehat{\bh}^H\bR_{\br} \widehat{\bh}}{\left\|\widehat{\bh}\right\|^2}\right]\\
& \stackrel{(c)}{\leq} \trace\left(\bR_{\widehat{\bh}}\right)+ \lambda_{\br,1}\\
& \leq \trace\left(\bR_{\widehat{\bh}}\right)+\lambda_1 \\
& \stackrel{(d)}{=} \sum_{t=1}^{T}\frac{\rho \lambda_t^2}{\rho\lambda_t +1} + \lambda_1
\end{align*}
where the inner expectation is over $\br$ (or noise $\bn$) and the outer expectation is over $\bh$ in $(a)$, $(b)$ comes from $\bw=\frac{\widehat{\bh}}{\left\|\widehat{\bh}\right\|}$, $(c)$ is because $\|\bR_{\br}\bx\|^2\leq \lambda_{\br,1}$ for any unit vector $\bx$, and $(d)$ can be easily derived similar to \eqref{tr_calc} with $\bX_{\mathrm{ss,opt}}=\sqrt{\rho}\bU_{[1:T]}$.
\qed

\begin{figure*}
\begin{align}
\nonumber  \bR_{1\mid 0} &=\eta^2 \bR_{0\mid 0} +(1-\eta^2)\bR \\
\nonumber    & = \bR - \eta^2 \bR\bX_{0,\mathrm{opt}}\left(\bI_T +\bX_{0,\mathrm{opt}}^H \bR \bX_{0,\mathrm{opt}}\right)^{-1}\bX_{0,\mathrm{opt}}^H\bR\\
\nonumber    & \stackrel{(a)}{=}\bU_0\Lambda_0\bU_0^H -\eta^2\left(\bU_0\mathbf{\Lambda}_{0[1:T]}\left(\bI_T+\rho\diag\left(\left[\lambda_{0,1},\cdots,\lambda_{0,T}\right]\right)\right)^{-1}
  \mathbf{\Lambda}_{0[1:T]}^H\bU_0^H\right)\\
\nonumber  & = \bU_0\diag\left(\left[\lambda_{0,1}-\eta^2\frac{\rho \lambda_{0,1}^2}{\rho \lambda_{0,1}+1},\cdots,\lambda_{0,T}-\eta^2\frac{\rho \lambda_{0,T}^2}{\rho \lambda_{0,T}+1},\lambda_{0,T+1},\cdots,\lambda_{0,N_t}\right]\right) \bU_0^H\\
  & = \bU_1\mathbf{\Lambda}_1\bU_1^H\label{R10}
\end{align}
\hrulefill
\end{figure*}
\begin{figure*}
\begin{align}
\nonumber \mathrm{MSE}\left(\bX_{1,\mathrm{opt}}\right) & = \frac{1}{N_t} \mathrm{tr}\left(\bR_{1\mid 0}-\bR_{1\mid 0} \bX_{1,\mathrm{opt}} \left(\bI_T + \bX_{1,\mathrm{opt}}^H\bR_{1 \mid 0} \bX_{1,\mathrm{opt}} \right)^{-1} \bX_{1,\mathrm{opt}}^H\bR_{1\mid 0}\right)\\
\nonumber & = \frac{1}{N_t} \mathrm{tr}\left(\bR_{1\mid 0}-\left(\bI_T + \bX_{1,\mathrm{opt}}^H\bR_{1 \mid 0} \bX_{1,\mathrm{opt}} \right)^{-1} \bX_{1,\mathrm{opt}}^H\bR_{1\mid 0}^2 \bX_{1,\mathrm{opt}} \right)\\
\nonumber &=\frac{1}{N_t}\left(\sum_{t=1}^{N_t}\lambda_{0,t}-\eta^2\sum_{t=1}^{T}\frac{\rho \lambda_{0,t}^2}{\rho \lambda_{0,t}+1}-\sum_{t=1}^{T}\frac{\rho \lambda_{1,t}^2}{\rho \lambda_{1,t}+1}\right)\\
&=1-\frac{1}{N_t}\left(\eta^2\sum_{t=1}^{T}\frac{\rho \lambda_{0,t}^2}{\rho \lambda_{0,t}+1}+\sum_{t=1}^{T}\frac{\rho \lambda_{1,t}^2}{\rho \lambda_{1,t}+1}\right)\label{mse_x1}
\end{align}
\hrulefill
\end{figure*}
\section{Proof of Lemma 4}
At $i=1$, $\bR_{1\mid 0}$ is given as in \eqref{R10} where $(a)$ comes from $\bX_{0,\mathrm{opt}}=\sqrt{\rho}\bU_{0[1:T]}$.  Note that $\bU_1$ and $\bU_0$ have the same columns with a different order based on the eigenvalues of $\mathbf{\Lambda}_1$.  Because $\bX_{1,\mathrm{opt}}=\bU_{1[1:T]}$, the MSE of the block $i=1$ is given as in \eqref{mse_x1}.  We can generalize \eqref{mse_x1} for $i>1$ with recursive derivation, which finishes the proof.
\qed

\bibliographystyle{IEEEtran}
\bibliography{refs}

\end{document}